**Predicting the risk of ischemic stroke in patients with atrial fibrillation using heterogeneous drug-protein-disease network-based deep learning**

Zhiheng Lyu, PhD[1,#], Jiannan Yang, PhD[2,#], Zhongzhi Xu, PhD[3,*], Weilan Wang, PhD[4], Weibin Cheng, MD[1,5], Kwok-Leung Tsui, PhD[6], Gary Tse, PhD[,7,8,9], Qingpeng Zhang, PhD[2,10,*]

1. School of Data Science, City University of Hong Kong, Hong Kong SAR, China.
2. Musketeers Foundation Institute of Data Science, The University of Hong Kong, Hong Kong SAR, China.
3. School of Public Health, Sun Yat-sen University, Guangzhou, China.
4. Centre for Healthy Longevity, Yong Loo Lin School of Medicine, National University of Singapore.
5. Institute for Healthcare Artificial Intelligence, Guangdong Second Provincial General Hospital, Guangzhou, China.
6. Grado Department of Industrial and Systems Engineering, Virginia Polytechnic Institute and State University, United States.
7. Tianjin Key Laboratory of Ionic-Molecular Function of Cardiovascular Disease, Department of Cardiology, Tianjin Institute of Cardiology, Second Hospital of Tianjin Medical University, Tianjin, China.
8. Cardiac Electrophysiology Unit, Cardiovascular Analytics Group, Kowloon, Hong Kong, China.
9. School of Nursing and Health Studies, Hong Kongetropolitan University, Homantin, Hong Kong, China.
10. Department of Pharmacology and Pharmacy, LKS Faculty of Medicine, The University of Hong Kong, Hong Kong SAR, China.

\# Equal contribution
\* Correspondence to
Zhongzhi Xu (xuzhzh26@mail.sysu.edu.cn)
Qingpeng Zhang (qpzhang@hku.hk)

**ABSTRACT**

Existing risk assessment models for predicting ischemic stroke (IS) in atrial fibrillation (AF) patients overlook the impacts of medications and the intricacy of drug-protein-disease interplays. We aim to develop an interpretable deep learning model, namely the **AF-Bio**logical-**IS-Path** (**ABioSPATH**), to predict the one-year IS risk in AF patients, and elucidate the underlying molecular mechanisms by integrating drug-protein-disease pathways and real-world clinical data of AF patients. A heterogeneous multilayer drug-protein-disease network was constructed to identify the mechanism of drug action and disease comorbidity propagation pathways. By integrating mechanism pathways and patient characteristics, ABioSPATH generated the IS risk and potential pathways for each patient. The Electronic Health Record (EHR) data of 7859 patients diagnosed with AF between 01/2008-12/2009 from 43 hospitals in Hong Kong were included. ABioSPATH outperformed all baselines across all metrics (AUROC=0.7815; 95% CI: 0.7346-0.8283, PPV=0.430, NPV=0.870, sensitivity=0.500, specificity=0.885, average precision=0.409, Brier score=0.195). In the cohort-level analysis, key proteins in the ten most common pathways, including CRP, REN, and PTGS2, were identified. A subsequent individual level analysis revealed the PIK3/Akt and cytokine and chemokine-induced signaling pathways among the top twenty important ones. Potential IS risks tied to less-studied drugs like prochlorperazine maleate were also highlighted. ABioSPATH demonstrated superior predictive performance in estimating IS risk with valuable molecular-level insights for clinical use. The advantage is that only routinely collected data are required, without requiring expensive biomarkers to be tested. Its applications extend beyond IS, indicating potential for risk screening in other diseases, enhancing patient care, and offering insights for drug development.

**Keywords**: Network Medicine; Network Science; AI in Medicine; Ischemic Stroke; Atrial Fibrillation

# 1 INTRODUCTION

Atrial fibrillation (AF) is the most prevalent cardiac arrhythmia, affecting 1% to 2% of the general population [1]. Ischemic stroke (IS) and systemic embolism are amongst the most ruinous complications of AF, irrespective of the types or patterns of AF. IS is one of the most fatal diseases in the world, with potentially deleterious outcomes such as vision and/or speech loss, paralysis, and loss of consciousness. Globally, there are 15 million people develop stroke each year, leading to 5 million deaths and 5 million permanent disabilities [2]. The burden of stroke is not limited to the patients and their families, it is also an issue for society. Oral anticoagulation therapy has been the pharmacologic standard for stroke risk reduction in patients with AF in both diagnostic and prognostic process but increases the risk of major bleeding. Fortunately, there is convincing evidence that IS is not inevitable in the course of aging if the risk can be predicted in time [3]. Extensive retrospective and prospective studies have unraveled a set of risk factors that can be adjusted to reduce the risk of IS [3, 4]. In clinical settings, $CHADS_2$ and $CHA_2DS_2$-VASc scoring systems are used for anticoagulation decision in AF patients [5]. However, they have been reported with ordinary performance, leading to potential overtreatment or undertreatment [6, 7]. Two gaps may attribute to the underdevelopment of predicting IS in patients with AF: (a) The co-occurrence of diseases can affect the risk of stroke through the underlying biological network of shared and multifunctional proteins and pathways [8]. Some of the major risk factors for stroke include high blood pressure, diabetes, heart and blood vessel diseases such as coronary heart disease, atrial fibrillation, heart valve disease, and carotid artery disease [9, 10], while the biological mechanism of such association is still unclear [11]; and (b) Drugs that the patient has taken may also influence the development of stroke in either positive or negative way. As an example, statins are found to be associated with a better prognosis outcome following IS with atrial fibrillation, although statins are usually prescribed as lipid lowering drugs [12]. This may be credited to the pleiotropic effects of statins on cardiovascular system independent from lipid control [13]. For example, studies show that drugs that target IL6 protein pathways may increase the risk of death due to IS [14], drugs that treat HIV have been associated with increased risk of IS [15]. Taken together, research that systematically harnesses heterogeneous drug-protein-disease interaction knowledge to improve our ability of IS risk prediction is lacking.

To fill such research gap, in this study we first constructed a multilayer network that integrates existing knowledge about disease-protein-drug multidimensional interactions. The model was developed using a territory-wide study cohort from Hong Kong, focusing on an Asian population. We then developed a deep learning model, namely the **AF-Bio**logical-I**S-Path** model (**ABioSPATH**) to predict the risk of IS in patients with AF and validated it in a Hong Kong-based cohort. Results showed that the proposed multilayer network-based knowledge framework significantly improved the performance of predicting the development of IS compared to baseline models, in which concurrent diseases as well as prescriptions, are considered independently. Moreover, ABioSPATH is able to generate explainable rules about pathways and proteomics that rationalize the risk prediction made by the model.

## 2 METHODS

### 2.1 Data

Data used in this study consisted of two parts. The first part was inpatients' Electronic Health Records (EHRs). They were collected from the Hong Kong Hospital Authority (HA), an institution that manages all 43 public hospitals in Hong Kong. For each admission, EHR recorded a unique patient identifier, gender, age, diagnosis (ICD-9-CM, up to 15), date of hospital admission, date of hospital discharge, and medications. Description date, British national formulary code, drug name, frequency, dosage value, dosage unit, description quantity, type of medication, and duration days were available for medication. Three years of EHR data, from Jan 1, 2008, through December 31, 2010, were collected and used in the analysis of this study. It contained over 5.2 million electronic health records (visits) covering 1,764,094 inpatients.

The second part of the data used in this study was a drug-protein-disease three-layer network harnessed from iDPath and the comorbidity network in our previous work as shown in Figure 1 [16, 17].

Among the three layers, the disease comorbidity network (DCN) illustrated the general knowledge that how diseases progress from one to another; The protein-protein interaction network (PPIN) revealed how proteins interact with each other,; There was no direct link in the drug network (DN), the connections between drugs and proteins were mediated through Drug-Protein Interactions . There were also interlayer links between DCN and PPIN, depicting the disease-protein relationships via The Disease-Protein Interactions. There were 13758 nodes, 1077 nodes and 27230 nodes in PPIN, DCN and DN respectively.

**Figure 1. A visualization of the drug-protein-disease network with 19 (out of 27230) nodes in the drug layer, 137 (out of 13758) nodes in the protein layer, and 53 (out of 1077) nodes in the disease layer.**

## 2.2 Sample selection

The sample selections is shown in Figure 2. ICD-9-CM code 427.31(**atrial fibrillation**) was used for identifying patients with AF, and 433(**Occlusion and stenosis of precerebral arteries**), 434(**occlusion of cerebral arteries**), 435(**Transient ischemic attack**), and 436(**Acute, but ill-defined, cerebrovascular disease**) for IS [18, 19]. During the study period, 7,859 patients were diagnosed with AF. Of the 7,859 diagnosed AF patients, 1,309 (25%) developed IS within 12 months. These patients were included in the analysis as positive cases. We included all the prescriptions before the first-time development of IS and all the diagnoses before the last diagnosis of AF. We acknowledge that the observed stroke rate in our study was relatively high when compared to other studies. However, there are several factors that might account for this discrepancy. Firstly, our model took into consideration recurrent strokes, which might increase the observed rate. Secondly, the age demographic of our study cohort leaned towards the older side, which inherently carries a higher risk for strokes. Additionally, we excluded patients for whom we lacked subsequent medical records. Many of these exclusions were due to patients being healthy and not experiencing a stroke in the year following their initial data entry. What's more, nearly 6000 patients were excluded from research due to a lack of corresponding prescription records. This could imply that the actual stroke rate might be lower than reported. Furthermore, our analysis also included patients who did not survive beyond one year. Taking all these factors into account provides a comprehensive explanation for the elevated stroke rate observed in our study.

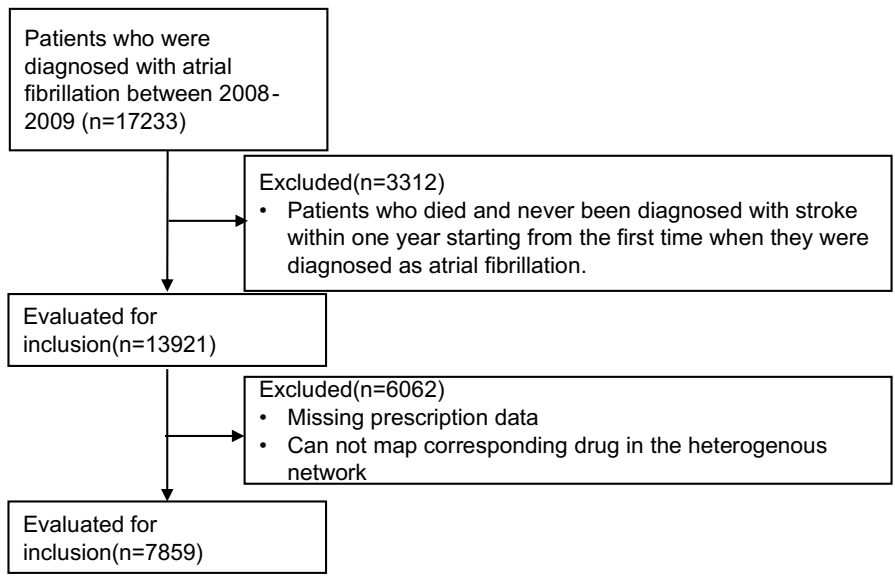

**Figure 2. Participants selection process.**

## 2.3 The ABioSPath model

Figure 3 illustrates the overall design of ABioSPATH, comprising two modules: a wide module and a deep module, which is inspired from the popular wide and deep model developed by Google [20]. The deep module takes a patient's prescription information and historical diagnoses as input and output the probability of developing IS within 12 months with a deep neural network structure. The basic assumption underlying the model is that the risk tends to propagate along the shortest path within the drug-disease-protein network, which has been proposed and tested in a few network medicine studies [21]. More concretely, given a patient's prescription information and historical diagnoses, the model first identifies the shortest path between the source nodes (i.e., historical drugs and diagnoses in the network) and the target node (i.e., IS). A Graph Attention Network (GAT) is used to capture local as well as global topological connectivity in a way of representing each node as a vector [22]. Then, the path vectors are fed into two Bidirectional Long Short-Term Memory (Bi-LSTM) neural networks [23]. Last, path embeddings are concatenated and fed into a three-layer multilayer perceptron to predict the "deep probability" of having IS in the next 12 months. The wide module uses a one-hot encoding technique to incorporate relevant information such as previous diagnoses, prescriptions, age, and gender. All previous diagnoses and prescriptions were represented as one-hot vectors. Specifically, diseases were encoded into a vector of length 555, drugs into a vector of length 109, age was categorized into three groups (below 65, between 65 and 75, and above 75) resulting in a vector of length 3, and gender was bifurcated into male and female, producing a vector of length 2. This culminated in a combined vector length of 669. This information is then fed into a single fully connected layer, which produced a "wide module probability" of an individual developing IS within the next 12 months. The final layer of the model combines the scores generated by both the wide and deep modules, resulting in a comprehensive prediction for an individual's risk of developing IS within the next 12 months. In this way, the model harnesses not only the patient's demographic and clinical information, but also the general knowledge about drug-protein-disease interaction information that is veiled in the multilayer network. We hypothesized that incorporating knowledge enrichment would enhance the accuracy of predicting an individual's risk of developing IS.

More importantly, the model is also designed to generate explanations about how the predicted result is made. Transparency is a critical aspect of any model, particularly in the healthcare industry, as it enhances the credibility and trustworthiness of the model among healthcare professionals and patients [21]. We introduce node attention layers and path attention layers to aggregate the node embeddings along the paths and generate the embedding of each path. The attention layers can (a) distinguish the association strength of nodes in each path, and (b) the contribution of each path to the final prediction.

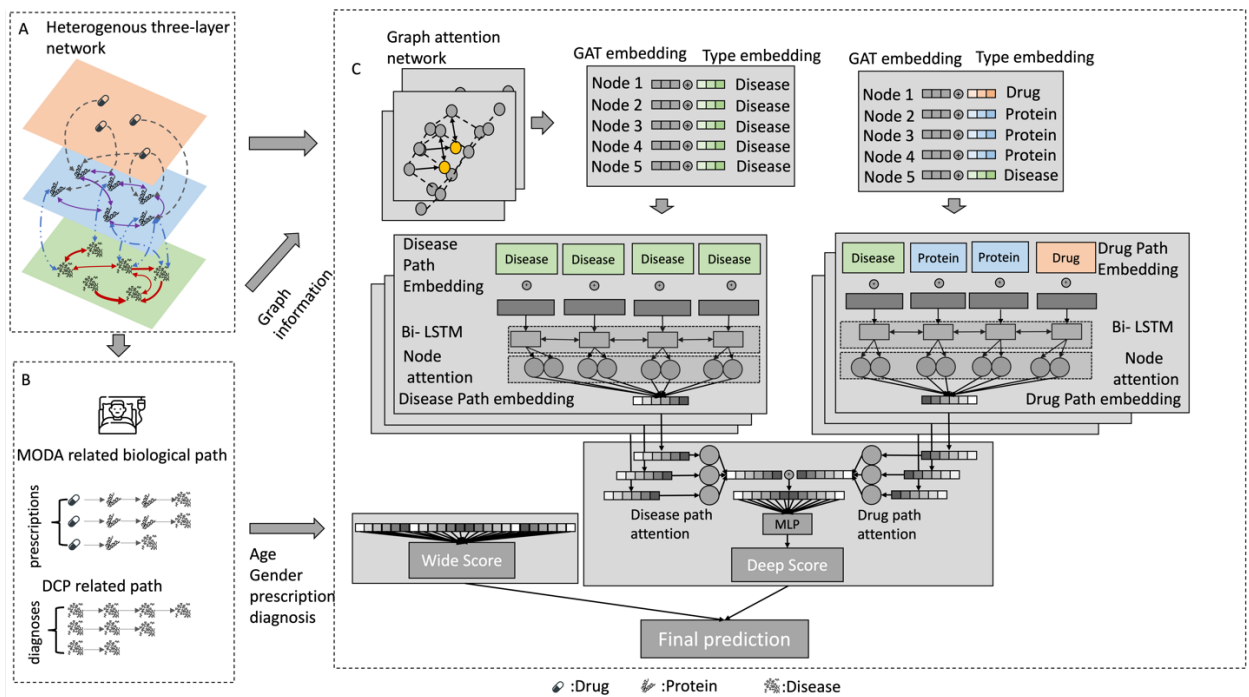

**Figure 3. The architecture of the proposed ABioSPATH model.** A. The heterogeneous biological network consists of four parts: two inter-networks and two intra-networks. More concretely, they are the drug-protein network (DN, in orange, with no intra-network links), the protein-protein interaction network (PPIN, in blue, with bidirectional purple arrows), the disease comorbidity network (DCN, in green, with bidirectional red arrows), and the disease-protein network. The dashed arrows from DN to PPIN represent relations between drugs and their targeted proteins; the dotted arrows between PPIN and DCN symbolize impacts between linked proteins and diseases. B. The mechanism of drug action (MODA)-related biological paths is identified using the shortest paths between drugs and ischemic strokes (IS) within this multilayer network. The disease comorbidity propagation (DCP) related paths are generated between past diagnoses and IS within the DCN. Age and gender are also integrated into the model. C. The algorithms' schematic representation shows how the heterogeneous biological network is fed into a three-layer Graph Attention Network (GAT) to learn the nodes' embeddings. The GAT embeddings for MODA paths and DCP paths are input into two separate Bidirectional Long Short-Term Memory (Bi-LSTM) networks to learn sequential relations. Node attention and path attention mechanisms are aggregated to learn nodes' and paths' embeddings, from which a deep score is generated. A one-hot encoding that includes past diagnoses, prescriptions, age, and sex information, computed based on the $CHADS_2$ score methodology, is fed into a
8

single-layer fully connected network to generate a wide score. A bias-free linear combination of the deep and wide scores yields the final risk score for each patient.

**2.4 Model Evaluation**

The research samples were randomly split into three parts. Seventy percent (5,503) of the study sample was used for model derivation, twenty percent (1,571) for model validation, and the remaining ten percent (785) for testing. A logistic regression model and least absolute shrinkage and selection operator (LASSO) using one-hot embedding together with clinical risk scores were adopted as baseline models [24, 25]. We used sensitivity, specificity, positive predictive values (PPV), negative predictive values (NPV), average precision and area under receiver operating characteristic curve (AUROC) metrics to evaluate the performance of all models. The logistic regression model was trained using 5-fold cross-validation on the training dataset. It employed the elastic net method for regularization with an optimal alpha value of 10-6 and an l1 ratio of 0. The model was then tested on both the validation and test datasets. Similarly, the LASSO model was trained using 5-fold cross-validation. It utilizes l1 norm regularization, and a parameter search determined the optimal alpha to be 0.045816. The LASSO model was also tested on the validation and test datasets. Both models use the same vector length, consistent with that of the wide module.

**3 RESULTS**

The characteristics of the research samples are shown in **Table 1**. The mean age of the study population was 79.82, ranging from 65 to 107. 56.5% were males. The average number of prescriptions for each patient during the study period was 4.11. The average number of diagnoses for each patient during the study period was 6.98.

**Table 1. The demographic and clinical characteristics of the participants**

| | |
|---|---|
| Mean Age (year) | 79.82 |
| Age range (year) | 65 – 107 |
| Sex: Male | 4439 (56.5%) |
| Average prescriptions used encoded by CID codes | 4.115 |
| Range of prescriptions used encoded by CID codes | 1 – 23 |
| Average diagnosis encoded by ICD9 codes | 6.984 |
| Range of diagnosis encoded by ICD9 codes | 1 - 35 |
| Number (percentage) of patients diagnosed with diabetes mellitus | 2175(27.68%) |
| Number (percentage) of patients diagnosed with essential hypertension | 4352(55.41%) |

## 3.1 Model performance.

The performance of the ABioSPATH model and baseline models are presented in Table 2. In general, ABioSPATH outperformed all baseline models in terms of all metrics. The calibration was assessed using a weighted Bier score. The AUROC, sensitivity, PPV, NPV, specificity, average precision and brier score were 0.7815, 0.430, 0.870, 0.5, 0.885, 0.409, 0.195 respectively. $CHADS_2$ and $CHA_2DS_2$-VASc achieved the worst result, followed by Logistic regression, LASSO. The relative statistical power of each baseline model in comparison to ABioSPATH was evaluated using the DeLong test.

**Table 2. Performance metrics.**

| Model | Cohort | AU-ROC | PPV | NPV | Sensitivity | Specificity | Average precision | Delong Test p-value | Brier-Score |
|---|---|---|---|---|---|---|---|---|---|
| Logistic regression | train | 0.745 (0.727, 0.762) | 0.449 | 0.8316 | 0.499 | 0.876 | 0.391 | <0.01 | 0.213 |
|  | valid | 0.735 (0.704, 0.766) | 0.4679 | 0.8517 | 0.500 | 0.887 | 0.398 | <0.05 | 0.213 |
|  | **test** | **0.725 (0.666, 0.782)** | **0.371** | **0.833** | **0.495** | **0.855** | **0.339** | **<0.05** | **0.221** |
| LASSO | train | 0.686 (0.669, 0.702) | 0.436 | 0.832 | 0.500 | 0.869 | 0.302 | <0.01 | 0.295 |
|  | valid | 0.698 (0.670, 0.727) | 0.457 | 0.844 | 0.500 | 0.870 | 0.322 | <0.01 | 0.284 |
|  | **test** | **0.648 (0.602, 0.695)** | **0.365** | **0.812** | **0.420** | **0.870** | **0.239** | **<0.01** | **0.308** |
| CHADS$_2$ | train | 0.667 (0.647, 0.685) | 0.316 | 0.741 | 0.366 | 0.840 | 0.262 | <0.01 | 0.231 |
|  | valid | 0.668 (0.633, 0.703) | 0.306 | 0.744 | 0.375 | 0.824 | 0.266 | <0.01 | 0.230 |
|  | **test** | **0.675 (0.625, 0.725)** | **0.256** | **0.734** | **0.322** | **0.837** | **0.231** | **<0.01** | **0.224** |
| CHA$_2$DS$_2$-VASc | train | 0.661 (0.642, 0.680) | 0.315 | 0.690 | 0.314 | 0.862 | 0.255 | <0.01 | 0.232 |
|  | valid | 0.658 (0.623, 0.693) | 0.313 | 0.701 | 0.327 | 0.852 | 0.255 | <0.01 | 0.232 |
|  | **test** | **0.673 (0.624, 0.722)** | **0.268** | **0.701** | **0.287** | **0.866** | **0.221** | **<0.01** | **0.220** |
| ABioSPATH | train | 0.813 (0.798, 0.827) | 0.500 | 0.843 | 0.500 | 0.900 | 0.451 |  | 0.183 |
|  | valid | 0.772 (0.737, 0.807) | 0.480 | 0.870 | 0.500 | 0.891 | 0.447 |  | 0.196 |
|  | **test** | **0.782 (0.735, 0.829)** | **0.430** | **0.870** | **0.500** | **0.885** | **0.409** |  | **0.195** |

## 3.2 Identifying residual risk in patients using certain medications

Given the constraints of our dataset, we lack data on oral anticoagulation drugs. However, we believe that aspirin and ACE inhibitors represent the most optimal medical therapies available in our dataset. Even when patients receive what's considered 'optimal' medical therapies, such as antiplatelet drugs and ACE inhibitors, there remains a persistent residual risk. We created a subset of patients who have been prescribed antiplatelet drugs (aspirin, clopidogrel, dipyridamole) and ACE inhibitors (lisinopril, ramipril, enalapril, captopril, fosinopril sodium, perindopril tertbutylamine), 6072 patients were included. Table 3 shows the performance of the ABioSPATH model and baseline models on the subset dataset. In identifying residual

IS risk within this subset, ABioSPATH outperformed other methods, showcasing an AUROC of 0.8067, PPV of 0.478, NPV of 0.787, sensitivity of 0.500, specificity of 0.897, average precision of 0.470 and Brier score of 0.186. In comparison, CHADS$_2$ and CHA$_2$DS$_2$-VASc fared the worst, followed by LASSO and then Logistic Regression.

**Table 3. Subset performance metrics.**

| Model | AUROC | PPV | NPV | Sensitivity | Specificity | Average Precision | Delong Test p-value | Brier score |
|---|---|---|---|---|---|---|---|---|
| Logistic regression | 0.748 (0.731,0.765) | 0.463 | 0.837 | 0.500 | 0.872 | 0.400 | <0.01 | 0.213 |
| LASSO | 0.692 (0.676,0.708) | 0.453 | 0.833 | 0.500 | 0.858 | 0.325 | <0.01 | 0.298 |
| CHADS$_2$ | 0.667 (0.650,0.685) | 0.327 | 0.733 | 0.383 | 0.824 | 0.279 | <0.01 | 0.233 |
| CHA$_2$DS$_2$-VASc | 0.661 (0.643,0.678) | 0.326 | 0.691 | 0.334 | 0.846 | 0.271 | <0.01 | 0.234 |
| ABioSPATH | **0.807 (0.794,0.822)** | **0.478** | **0.878** | **0.500** | **0.897** | **0.470** | | **0.186** |

### 3.3 Identified Pathway results for drugs.

ABioSPATH incorporated the widely adapted deep learning way of understanding the mechanism of drug actions (MODA) and disease comorbidity propagations (DCP) into disease risk prediction.[16]

Amongst the identified pathways in all patients, we selected the ten most important pathways for each patient based on their ranked weights. We aggregated the selected pathways from all patients and presented the ten most frequently occurring ones as illustrated in Figure 4A. Among all twenty pathways, four started from CID-2244(aspirin), three started from CID-5362119 (lisinopril), two started from CID-3440 (furosemide), one started from CID-4510 (nitroglycerin). The identified proteins included Entrez-1636 (ACE), Entrez-5743 (PTGS2), Entrez-59272 (ACE2), Entrez-476 (RNA5SP476), Entrez-5972 (REN), Entrez-4843 (NOS2), Entrez-1559 (CYP2C9), Entrez-1401 (CRP), Entrez-4311 (MME). We illustrated an individual patient's identified drug pathway as an example in Figure 4B. This patient had a medical history of ICD9-250 (Diabetes mellitus), ICD9-272 (Disorders of lipid metabolism), ICD9-401 (Essential hypertension), ICD9-427 (Cardiac dysrhythmias), ICD9-428 (Heart failure), ICD9-438 (Late effects of cerebrovascular disease), ICD9-496 (Chronic airway obstruction, not elsewhere classified), and took CID-4917 (prochlorperazine maleate), CID-3746 (ipratropium bromide), CID-3108 (dipyridamole), CID-2244 (aspirin), CID-2153(theophylline), CID-2083 (salbutamol (sulphate)), CID-5362119 (lisinopril), CID-38853 (methyldopa), CID-5754 (hydrocortisone), CID-39186 (diltiazem), and later diagnosed with ischemic stroke within one-year follow-up. The twenty most important paths from drugs to strokes could be picked based on ranked weights. The specific pathways are shown in figure 4B. Among all twenty pathways, eight started from CID-2153 (theophylline), five started from CID-4917 (prochlorperazine maleate), three started from CID-2083 (salbutamol), two started from CID-38853 (methyldopa), one started from CID-2244 (aspirin), one started from CID-39186 (diltiazem). The longest length of pathways was four, the shortest was four. Fifty-two proteins appeared in those paths, the frequency of which ranged from one to five, the five most important proteins were Entrez-5290 (PIK3CA), Entrez-7124 (TNF), Entrez-1129 (CHRM2), Entrez-4804 (NGFR), Entrez-153 (ADRB1).

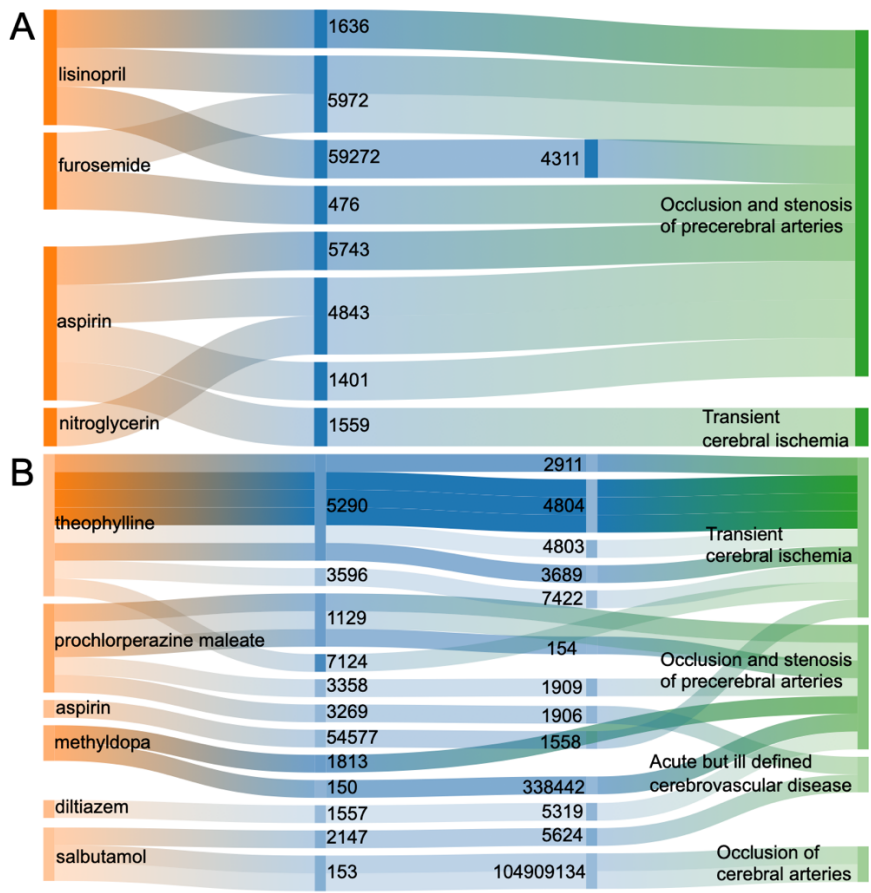

**Figure 4. (A) The Sankey diagram shows the top ten most common protein pathways linking medications to ischemic stroke across the study cohort. (B) The Sankey diagram shows protein pathways connecting medications and ischemic stroke in an example patient.** The size of the graph represents the frequency of the protein appearing in the pathways. Orange nodes represent drugs, blue nodes represent protein, and green nodes represent diseases. Color's darkness shows its association strength. Darker node colors indicate higher node association strength, while darker link colors signify greater link.

## 4 DISCUSSION

In this study, we initially constructed a multilayer network that integrates existing knowledge about disease-protein-drug multidimensional interactions. We subsequently developed a deep learning model, namely the ABioSPATH, to make use of the resulting multilayer network to improve the prediction capability of the risk of IS for people with AF. Experiments on a cohort of 7859 atrial fibrillation inpatients showed that ABioSPATH demonstrated an improvement in the AUROC by 8% when compared to the standard logistic regression model, which considered prescription and historical diagnoses independently.

In subset comparisons, our model excels over existing standards. Current clinical risk score standards overlook drug usage while ABioSPATH incorporates it. It accurately identifies high-risk patients, even under treatment deemed optimal by current standards. Consequently, our model effectively pinpoints patients with residual risk. Additionally, ABioSPATH provided valuable insights into the potential mechanical pathways that contribute to the prediction performance of the model. The high negative predictive value (NPV) further demonstrates that our model is highly accurate in identifying patients who do not have stroke risk. This capability can better guide clinicians in recognizing patients who are not at immediate danger of ischemic stroke, potentially helping to avoid unnecessary medication treatments and associated risks of internal bleeding.

The rationale of the predicted result is important in the medical and healthcare context because it can help doctors make informed decisions about patient care. Our model tracks the shortest paths in a complex, multi-layered network, originating from specific drugs and terminating at nodes representing ischemic stroke. We employed the cuGraph shortest path search function—a GPU-accelerated graph analytics package—to expedite the process given the network's extensive size [26]. ABioSPATH provides possible pathways that rationalize risk prediction. Protein-protein interaction plays a key role in predicting the target protein functions and drug impact at the molecular level. In drug discovery, protein-protein interactions are increasingly important, offering great potential for therapeutic modulation of target classes in disease. Compared with baseline models, our model excels in distinguishing critical paths from drugs to diseases.

Figure 4A demonstrates that the drugs aspirin, lisinopril, furosemide, and nitroglycerin significantly impacted a patient's IS risk more than other medications according to our model.

Lisinopril, an angiotensin-converting enzyme (ACE) inhibitor, promotes vascular relaxation and dilation, consequently lowering blood pressure [27]. Extensive research has elucidated the connection between ACE inhibitors and vascular diseases, specifically identifying the relationship with ischemic stroke [28]. Furthermore, REN plays a key role in the renin-angiotensin system, which regulates blood pressure and fluid balance [29].

Aspirin is widely used to prevent recurrent IS [30]. Numerous studies have suggested that PTGS2 may serve as a critical risk factor in IS, with genetic variation in PTGS2 potentially contributing to the development of cardiovascular events [31]. The utilization of aspirin seems to modify the relationship between the PTGS2 G-765C polymorphism and coronary heart disease (CHD) risk [32]. PTGS2 G765C mutation may be related to aspirin resistance [33]. Additionally, aspirin can influence the activity of nitric oxide synthase (NOS), a family of enzymes that includes NOS2 [33]. NOS2 has been identified as stroke related gene [34]. CRP has been identified as an important risk factor for ischemic stroke [35]. In atrial fibrillation patients, CRP levels correlate with stroke risk and recognized risk factors. CRP also links to other cardiovascular diseases and prognosis, serving as a key biomarker in risk assessment for IS [35-37]. Aspirin therapeutic efficacy has been observed among individuals with high CRP levels [35]. Evidence also supports the connection between furosemide, nitroglycerin, and IS via identified pathways [38, 39].

For the specific case, Figure 4B shows that among the top 20 most significant paths, the drugs theophylline and prochlorperazine maleate had the most substantial impact on the patient's stroke risk. The relationship between theophylline and IS is complex and multifaceted. Our pathway prediction indicates that theophylline is associated with ischemic stroke through proteins phosphatidylinositol-4,5-bisphosphate 3-kinase catalytic subunit alpha (PIK3CA) and Nerve Growth Factor Receptor (NGFR). Theophylline is commonly used in asthma and chronic obstructive pulmonary disease treatment [40], the relationship between ischemic stroke and theophylline remains under-explored. However, some evidence suggests that theophylline may have a neuroprotective effect in ischemic stroke [41]. PIK3CA is an isoform of the catalytic subunit of Phosphoinositide 3-kinases (PI3Ks), a key part in PI3K/Akt pathway and can be inhibited by theophylline [42-44]. PI3K/Akt signaling pathway has been shown to be involved in various disease pathogenesis including stroke [45]. NGFR is related and involved in the development of ischemic stroke, one study shows that NGFR might be nominated as a risk factor for developing ischemic stroke [46]. TNF has been identified as a potential marker for stroke due to its significant role in the pathophysiology of stroke and its involvement in cytokine- and chemokine-induced signaling pathways in neuroinflammation[45, 47]. There is proven evidence that other drugs identified in this study were found to be associated with IS. For example, aspirin was commonly used to prevent blood clot, the leading cause of IS. The administration of aspirin significantly lowered blood clot density and strength [48]. Similarly, methyldopa may also decrease the IS risk through its anti-hypertension effect [49]. Another drug associated with IS risk is prochlorperazine maleate. Despite current research has yet to definitively explore the relationship between prochlorperazine maleate and IS, there have been clinical case report suggesting a connection [50]. Notably, our model's result in line with the finding. Our model suggests that administering prochlorperazine maleate might increase the ischemic stroke risk in this patient. This conclusion comes from a rise in risk score from 0.559 to 0.566 post drug administration. Thus, our model can project the potential risks tied to the use of less-studied IS-related drugs, such as prochlorperazine maleate. Although pathways associated with IS risk raised from the ABioSPATH were identified by the algorithm instead of clinical practice, these pathways aligned with previous studies. Therefore, some under-explored pathways may be the new target in IS prevention.

Cohort level analyses provide statistical interpretations of overall pathways and relationships from drug to disease. In contrast, individual level analysis presents unique drug interactions and relations; thus, different drugs may have varying influences on specific diseases. This distinction underscores the strength of our model in integrating complex relationships and providing personalized results, representing a significant advancement.

The contribution of this study is two-fold. Firstly, to the best of the authors' knowledge, we have, for the first time proposed a dataset that established connections between links together the drug-protein and disease-protein interactions, as well as protein-protein and disease-disease inner relationships. With the collection of worldwide health medical data and high throughput interactome assay data being electronically, the quantitative characterization of disease co-occurrences and proteomics is gaining momentum. However, those data are isolated and not yet systematically harnessed. The dataset herein this study provides a foundation that facilitate not only the prediction of IS, but also the development of other diseases, while serving as a substrate for inspiring novel molecular pathways in disease development. Secondly, we developed a GAT-based deep learning model that incorporates the resulting multilayer network and validated it using large-scale real-life inpatient data. This model offers a valuable framework by linking the resulting network to patient data involving multiple comorbidities and prescriptions, which often induce difficulties in downstream analysis. In our results, we demonstrate that our model identifies significant proteins and pathways at both the cohort and individual levels. Our model adapts the results based on drug-protein-disease associations. In the cohort level analysis, proteins such as CRP, REN, and PTGS2 were identified as key elements associated with IS. At the individual level, two critical protein pathways, the PI3K/Akt signaling pathway and the Cytokine- and chemokine-induced signaling pathways were noted to be crucial. There findings were found align with those from other IS research studies.

The study has several important limitations that need to be acknowledged. Firstly, our analysis was conducted solely on patients from Hong Kong, which may restrict the generalizability of our findings to other parts of the world. It is crucial to validate our model with patients from diverse geographic regions to uncover potential variations in circumstances and ensure the broader applicability of our results. Secondly, we did not account for the temporal order of diseases and prescriptions in our study. While incorporating the chronological sequence of disease onset and corresponding prescriptions can introduce complexities and additional noise, it is essential to explore this aspect. By considering the temporal dimension, we could potentially discover valuable insights, patterns, and relationships that might otherwise remain hidden. Future research should aim to incorporate this temporal information to gain a more comprehensive understanding of the data. Thirdly, the weighting of the knowledge graph used in our study is a complex issue that warrants further investigation. Currently, we assigned fixed weights to the connections between drugs and proteins, without considering their relative association strength. However, these weights hold valuable information, representing the strength of the link between entities. To enhance the accuracy and richness of our model, our next focus of study will be on assigning weights based on the significance and relevance of different chemicals. This refinement will allow us to incorporate more valuable information into the knowledge graph and improve the overall performance of our model.

## 5 CONCLUSION

This study established a dataset connecting the drug-protein, disease-protein, as well as protein-protein and comorbidity relationships. Based on this dataset, we further constructed a multilayer network and developed a deep learning model, ABioSPATH, to predict the risk of ischemic stroke among patients with atrial fibrillation. ABioSPATH demonstrated a superior prediction performance compared to other machine learning methods and clinical risk scores, even among subgroups of patients receiving specific therapies. It also successfully identifies the underlying proteins and pathways from drugs to diseases, providing insights at the molecular level. The advantage is that only routinely collected data are required, without requiring expensive biomarkers to be tested. The implications of this work extend beyond the IS, this model shows potential for screening risks of other diseases and behavioral disorders. This study may serve as a steppingstone towards improved patient care, as well as laying the groundwork for personalized drug recommendation and patient care.

## DATA AVAILABILITY

Data used in this study are available upon request.

## CODE AVAILABILITY

Python code for constructing the ABioSPATH is available on GitHub

https://github.com/lyuzhathk/abiospath_lyu

## ETHICAL APPROVAL

This study was approved by the NTEC/CUHK Research Ethics Committee (reference no. 2018.643).

## CONSENT TO PARTICIPATE (HUMAN SUBJECTS)

This study is a retrospective secondary data analysis of anonymous electronic health record data. There is no consent to participate.

## COMPETING INTERESTS

The Authors declare no Competing Financial or Non-Financial Interests.

## AUTHOR CONTRIBUTIONS

Z.L.: Methodology, Validation, Formal analysis, Visualization, Writing - Original Draft.

J.Y.: Methodology, Validation, Formal analysis, Writing - Review & Editing, Project administration.

Z.X.: Methodology, Validation, Formal analysis, Writing - Review & Editing, Project administration.

W.L.: Formal analysis, Writing - Review & Editing.

W.W.: Formal analysis, Writing - Review & Editing.

K.L.T.: Formal analysis, Writing - Review & Editing.

G.T.: Methodology, Writing - Review & Editing, Project administration.

Q.Z.: Conceptualization, Methodology, Writing - Original Draft, Writing - Review & Editing, Supervision, Project administration, Funding acquisition.